# MAGNETIC SYSTEM FOR CLEANING THE GAMMA BEAM AT THE LUE-40 ELECTRON LINAC OUTPUT

*V.V. Mytrochenko, S.O. Perezhogin, L.I. Selivanov, V.Ph. Zhyglo, A.N. Vodin,*

*O.S. Deiev, S.M. Olejnik, I.S. Timchenko and V.A. Kushnir*

*NSC KIPT, Kharkiv, Ukraine*

*E-mail: mitvic@kipt.kharkov.ua*

The bremsstrahlung of accelerated electrons passing through a converter is used to study multiparticle photonuclear reactions. The results of calculations, numerical modeling, design, and testing of a special magnetic cleaning system to obtain a "pure" beam of bremsstrahlung quanta when studying the cross-sections of such reactions at the LUE-40 linac are presented. The system is based on commercially available permanent magnets of rectangular cross-sections. The maximum on-axis field is 0.9 T, which provides sufficient separation of the electron beam and gamma rays at a distance of more than 90 mm from the magnet.

## 1. INTRODUCTION

Cross-sections of the photonuclear reactions were mainly obtained experimentally at a study of the giant dipole resonance (GDR) using bremsstrahlung radiation and quasi monoenergetic photons [1-5] at γ- quanta energies up to 30 MeV. Recently obtained experimental results on the cross-sections of multiparticle photonuclear reactions at γ - quanta energies of 30–145 MeV are quantitatively many times different from the data on (γ,n) reactions.

Study of photofission of nuclei in the energy range above the GDR through to the threshold of pion production ($E_{th} \approx 145$ MeV) is of particular interest because there is a change in the mechanism of interaction of photons with nuclei in this energy region. Therefore it is possible to obtain fundamental data on two mechanisms of photofission of the nucleus, namely, due to GDR excitation and quasi-deuteron photoabsorption [6]. To date, various theoretical models, for example, [7,8], as well as the modern nuclear reaction codes [9-12], have been developed to describe photonuclear reactions at their cross-section calculations. Both the theoretical models and the calculation codes need new data to verify multiparticle reactions in a wide range of atomic masses and energies.

In practical applications, knowledge of the exact parameters of multiparticle photonuclear reactions is important, for example, to estimate the neutron yield in nuclear installations based on accelerator-driven subcritical systems (ADS). The ADS can be used for the utilization of radioactive waste of nuclear energy plants (transmutation of minor actinides [13] and burning long-lived fission fragments). The ADS are also potential sources for electricity generation [14,15].

The study of multiparticle photonuclear reactions with relatively small cross-sections (~0.1–100 mb) is possible in the presence of intense fluxes of the incident - γ quanta. Such fluxes can be obtained by passing beams of high-energy electrons through a target converter. Experiments with the use of bremsstrahlung, which has a continuous spectrum of the form ~ $1/E_\gamma$ and the maximum limiting energy $E_{\gamma max}$, make it possible to determine the integral characteristics of the reactions, such as the averaged yield $Y(E_{\gamma max})$ and average cross-sections $\langle\sigma(E_{\gamma max})\rangle$ of reactions. On the other hand, the application of such γ spectrum complicates the procedure for determining the

$\sigma(E_\gamma)$ cross-sections of photonuclear reactions because it requires additional calculations and the use of various data processing methods. In both cases there is a need for correct calculations of the γ-quanta flux density, which corresponds to the real conditions of the experiments, with modern computational codes, for example, such as GEANT4 [16]. However, despite the difficulties, bremsstrahlung beams remain an important tool in nuclear physics research [17-19].

In the experiments with bremsstrahlung, three main schemes can be distinguished. The first one uses a flux consisting of bremsstrahlung and electrons that have passed through the converter [17] that interacts with atoms of the target substance. The advantages of this method are mainly in simplicity of the approach, however high radiation load on the target leads to technical difficulties (sintering of samples, burning of structural elements of the target node, etc.). It is also necessary to consider a contribution to the reaction yield from the processes occurring under the action of electrons. Usually, this contribution can be calculated, although its probability is strongly suppressed by the mechanism of interaction of electrons through virtual photons. The constant of this interaction is proportional to the constant of fine structure which is about 1/137. However, for high-threshold reactions, the contribution of electrons may dominate. Thus, it was shown in [20] on the example of $^{93}$Nb(γ,xn)$^{93-x}$Nb reactions, where x is neutron multiplicity, that for the case of x=5 the yield of the reaction under the action of gamma quanta becomes less than that from the contribution of electrons.

The second scheme is similar to the first one, but after the converter, a massive electron absorber consisting of light material (usually Al) [21,22] is installed. The advantage of such a scheme is simplicity and almost a "pure" gamma quantum beam on the target. This guarantees a slight radiation and heat load of the target and, for example, on the components of the pneumatic tube transport that is used for target delivery. The disadvantages of this scheme are the distortion of the shape of the bremsstrahlung spectrum, the additional generation of photoneutrons, which also contribute to studied reaction yield.

The third scheme of the experiment implements a bending magnet to divert the electrons that have passed through the converter [22] that allows obtaining a "pure" beam of bremsstrahlung gamma quanta at the target. When using thin converters, the shape of the radiation spectrum can be described by a known analytical formula. At the same time, the influence of neutrons is minimized.

Thus, the study of photonuclear reactions with the "pure" gamma beam using the third scheme has the following advantages:

• Thin converter can be used because electrons will not create a radiation load and gamma rays will not scatter in the absorber. Accordingly, the bremsstrahlung spectrum can be described by an analytical formula of the form $\sim 1/E_\gamma$.

• The contribution of neutrons to the reaction outputs can be neglected.

• The possibility of using pneumatic tube transport remains.

• There is the possibility to obtain cross-sections of photonuclear reactions $\sigma(E_\gamma)$ with the photon difference method or the regularization method.

• Due to lack of the absorber, it is possible to significantly decrease the volume of calculations of bremsstrahlung quantum flax with the computer codes. Experimental and calculated errors are significantly reduced.

All the above mentioned makes the task of obtaining a "pure" beam of gamma-ray quanta for studies of cross-sections of multiparticle photonuclear reactions at the output of the LUE-40 accelerator [23] using a magnetic cleaning system relevant and will provide a set of new experimental data.

When developing such a cleaning magnetic system, it should be considered that the electrons are scattered in the converter, so the system should be short, with significant induction of the magnetic field in the interaction space to avoid electron beam collimation by magnet poles. Estimates have shown that magnetic induction value should be around one Tesla. The most suitable magnetic system for this purpose is a system based on permanent magnets with a special direction of magnetization [24]. One of the implementations of such a magnet is given in the examples to the Superfish/Poisson group of codes [25]. However, such a system requires permanent magnets with a special configuration, which are not widely available. Therefore, commercially available magnets with a rectangular cross-section were used for the prototype of the magnetic system.

The paper presents the results of calculations, numerical simulations, design, and testing of such a system to obtain a "pure" beam of bremsstrahlung quanta for the study of cross-sections of multiparticle photonuclear reactions at the LUE-40 electron linac.

**2. MAGNETIC SYSTEM SIMULATION**

Simulation of transverse magnetic field distribution was performed with the two-dimensional Superfish/Poisson group of codes under the assumption that the magnetic system extends infinitely in the longitudinal direction. For such an approach to be applied to a finite-sized magnet, the air gap of the magnet must be much smaller than the longitudinal length of the system. Based on the geometric dimensions and mutual positioning of the magnet and the irradiation target, it is necessary, that the bending angle of the beam central trajectory at the energy of 80 MeV was not less than 15°. In the physics of accelerators, a quantity such as magnetic rigidity is used (see, for example, [26]): $B\rho = 3.3356 pc$, where $B$ is the magnetic field in Tesla, $\rho$ is the electron Larmor radius in meters, $p \cdot c$ is electron energy in GeV. It can be shown that for a rectangular magnet at small bending angles $\alpha$ the length of the magnet $L$ is determined by the formula: $L \approx 3.3356 pc\alpha / B$. At the air gap magnetic field of 0.9 T, the length of the magnet at an electron energy of 82 MeV is approximately 80 mm. NdFeB magnets with a size of 80x40x15 mm$^3$ are commercially available, so with an air gap of 20 mm or less, a two-dimensional code can be used to simulate the magnet.

To increase the field in the gap, a magnetic yoke made of low-carbon steel was used, which closed the magnetic flux, forming a so-called "C"-like magnet. The following magnet parameters were used in the simulation: residual induction $B_r$ = 12100 Gauss and coercive force $H_c$ = -11300 Oersted. This corresponds to the N38 class NdFeB magnets we purchased.

A magnetic system simulation of certain optimization was performed with such competing criteria as field induction in the gap, the size of the gap, the maximum induction in the magnetic yoke. The minimum size of the gap was limited by the electron beam size considering scattering in the converter. The maximum induction in the gap should be approximately 0.9 T, the maximum induction in the magnetic yoke should not exceed 1.5 T.

The final configuration of the magnetic system (it's upper half relative to the plane of symmetry) is shown in Fig. 1.

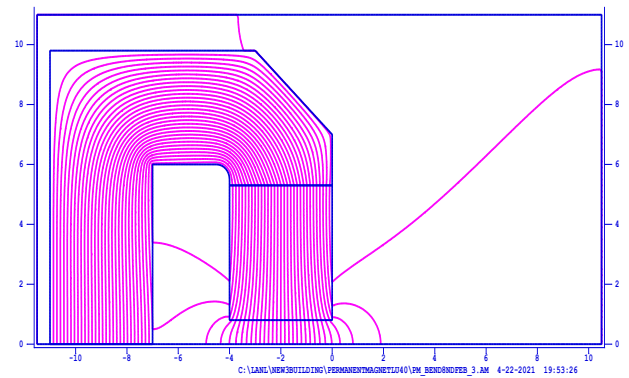

*Fig. 1. Magnet system configuration.*

The system shown in Fig. 1 consists of a block of permanent magnets (three plates with a total thickness of 45 mm) and the "C" like magnetic yoke. The thickness of the vertical part of the magnetic yoke is 40 mm, and that of the horizontal part is 45 mm. To reduce the magnitude of the field induction at the point of contact of the permanent magnets with the magnetic yoke on the left, a notch of 7 mm deep with a rounding radius of 5 mm was made in the horizontal part of the magnetic yoke. With a magnetic gap of 16 mm, the induction in the gap is 0.876 T, and this value in the magnetic yoke does not exceed 1.5 T.

Fig. 2 shows the calculation results of the central trajectory for the case of an electron beam with energies of 44.3 MeV and 82 MeV using the TRANSPORT code [27]. You can see that the magnet provides a beam deflection of 8 cm and 4 cm at the distance 10 cm from the magnet, respectively.

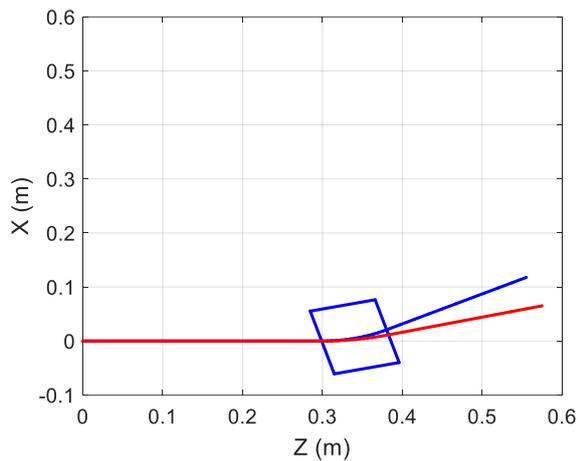

Fig. 2. The central trajectories of the electron beam for energies of 44.3 MeV (blue curve) and 82 MeV (red curve).

### 2.1 SIMULATION WITH PARMELA and GEANT4.

An array of macroparticles, which represents an electron beam simulated by the PARMELA code [28] at the LUE-40 linac output was introduced into the GEANT4 program as an event generator. This electron beam fell on a tantalum converter 0.1 mm or 0.3 mm thick. The energy of the beam macroparticles at the maximum of the distribution was 44.3 MeV, 70% of the particles fell into the energy range of 3.2%. The number of macroparticles in the beam was 190,700. The parameters of the electrons that flew forward from the converter were transferred back to the PARMELA code for further transport of macroparticles. There were two sets of simulations. At the first set beam divergence after converter was studied, so beamline was represented just a 10 cm drift space. At a converter thickness of 0.1 mm, the RMS beam size increases in the drift space by 0.56 cm, which corresponds to a beam standard deviation of 56 mrad. With a converter thickness of 0.3 mm, this increase equals to 1 cm, which corresponds to the beam standard deviation of 100 mrad. It should be noted that the number of macroparticles transmitted to the PARMELA program is greater than their initial number. At a converter thickness of 0.1 mm, they are approximately 3,000 particles larger, and at a thickness of 0.3 mm that increase is approximately 5,000 particles.

At the second set, the effect of the magnet was studied. The beamline was represented by a 0.5 cm drift, the dipole magnet, and several drifts with a total length of 10 cm. For the energy of the reference particle 44.3 MeV and the magnetic field induction of 0.9 T, the bending angle of the magnet was 27.5°. The magnet was set at an angle of 13.8° to the axis of the input beam. At a converter thickness of 0.1 mm, 92% of particles pass through the magnet, and at a thickness of 0.3 mm that value was 70%. The horizontal beam profiles (bending plane of the magnet) for different thicknesses of the converter are shown in Fig. 3. These profiles look like the Gaussian distribution. One can see that for the converter with a thickness of 0.1 mm there is a very small number of particles at 3 cm from the beam central trajectory. For the converter with thickness of 0.3 mm, approximately the same number of particles is observed at 5 cm.

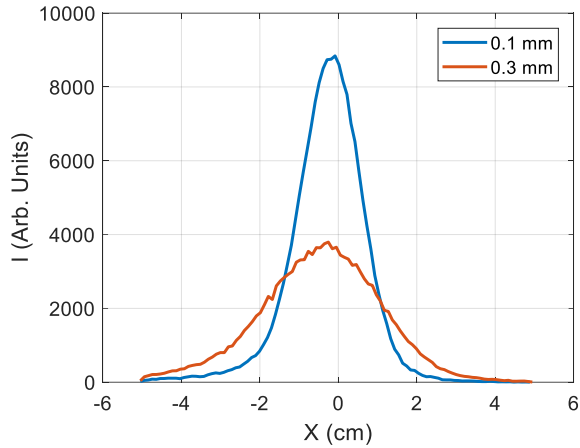

*Fig. 3. Horizontal profiles of the electron beam at 10 cm from the magnet output edge. The blue curve corresponds to a converter thickness of 0.1 mm and the red curve corresponds to the thickness of 0.3 mm*

A slightly different picture is observed in the vertical plane (see Fig. 4). It is seen that there are no "tails" in the distributions. Thus, at a converter thickness of 0.1 mm, the vertical profile of the beam shows signs of limitation by the aperture of the magnet at a level of approximately 10% of the maximum. However, with a converter thickness of 0.3 mm, this limitation occurs at a level of approximately 40%. This explains where particles that do not pass through the magnet are lost.

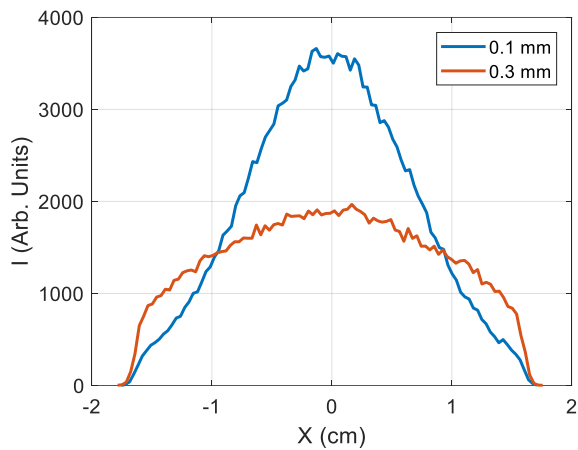

*Fig. 4. Vertical profiles of the electron beam at 10 cm from the magnet output edge. The blue curve corresponds to a converter thickness of 0.1 mm and the red curve corresponds to the thickness of 0.3 mm*

Therefore, based on the above simulation results, it can be concluded that such a magnet design is suitable for cleaning the gamma ray beam from electrons, at least for a converter thickness of 0.1 mm. It is known that the scattering angle of relativistic electrons is inversely proportional to their energy, therefore, at particle energies in the region of 80 MeV, a tantalum converter with a thickness of 0.3 mm can be used. This question requires a more detailed study to assess the effect of background radiation from the collision of electrons with a magnet on the results of multiparticle photonuclear reactions in the target.

## 3. MANUFACTURE AND TESTING OF MAGNETIC SYSTEM

The developed magnet was manufactured by Science and Research Establishment "Accelerator" with NSC KIPT. The magnetic yoke consists of three parts fastened with steel bolts. These parts are made of low-carbon steel. Purchased magnetic plates were assembled in two blocks of three plates. A screw device that allows one to hold the magnets when pushing them on top of each other was used at assembling the blocks and installing the blocks in the magnetic yoke. Magnets in blocks and blocks in a magnetic yoke are fixed with an aluminum framework. The appearance of the magnet is shown in Fig. 5.

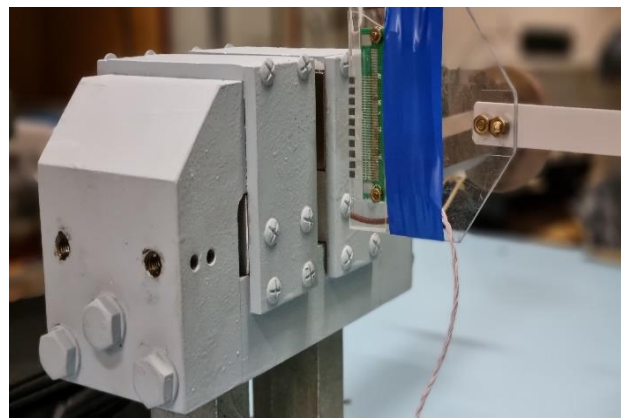

*Fig. 5. The magnet with a line of Hall sensors for measuring the magnetic field distribution.*

To measure the field distribution in the magnet, a line of 10 linear Hall sensors CYSJ902 [29] was made, which allows measuring fields up to 2 Tesla with a relative error of up to 1%. To calibrate the sensors, each of them in turn was placed at a point with the same field. The field at this point was measured with an industrial teslameter. The measurement results are shown in Fig. 6. The maximum induction in the gap was 0.9 Tesla.

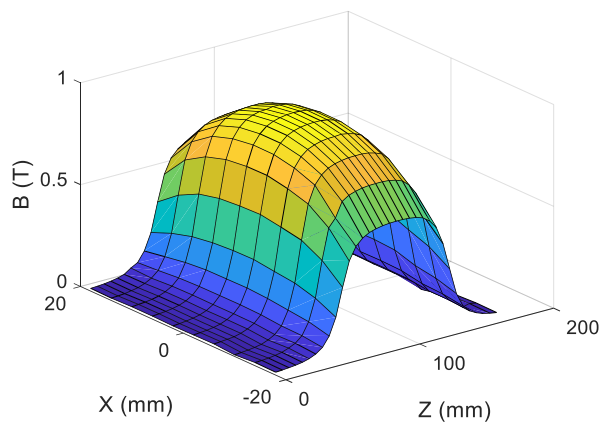

Fig. 6. Magnetic field distribution in a midplane of the magnet air gap.

A comparison of the measured data across the magnet in the middle of the pole with the dependence calculated using the Superfish/Poisson code was performed. The measured data were normalized so that the maximum value coincided with the calculated maximum value. The measured field distribution in shape coincides well with the calculated one. In absolute terms, the results differ by 3%.

The integration of the field distribution in the middle of the magnet poles in the longitudinal direction allowed us to estimate the fringing fields. The effective length of the magnet is 82.7 mm with a pole length of 80 mm, which indicates that the field is well localized within the magnet.

The magnet was installed at the LUE-40 linac output (see Fig. 7) at an angle of 15° to the linac axis. To study the effect of the magnet on electron beam dynamics, a series of experiments was performed at beam energies in the regions of 40 MeV and 80 MeV. It is known that electron irradiation causes silicon glass coloring (see, for example, [30]) that depends on the absorbed dose, so it is the simple way to visualize beam transversal density distribution. A glass plate was installed at 90 mm from the magnet exit in such a way that the plate plane was perpendicular to the linac axis at every experiment including a configuration without any converter as well as the configuration with installed 0.1 mm and 0.3 mm thick tantalum converters.

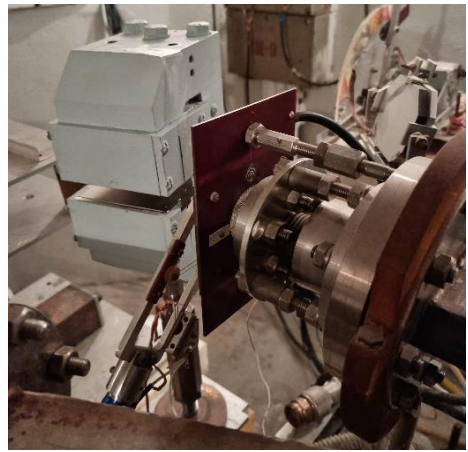

Fig. 7. The cleaning magnet was installed at the LUE-40 linac exit.

An example of photographs of beam footprints on the checkered paper background is shown in Fig. 8 at particle energy of 81 MeV. The black lines at the bottom right correspond to the horizontal position of the linac axis. A dash at the bottom of the lower footprint is accidental contamination of the glass.

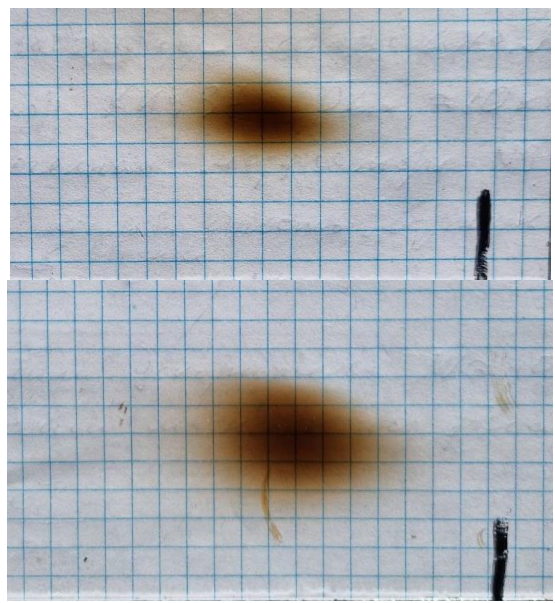

Fig. 8. Photographs of beam footprints on glass at energy of 81 MeV. Top - without converter, bottom - with 0.1 mm tantalum converter.

It can be concluded that the developed magnet, in principle, can be used to clean the gamma beam at an electron beam energy of up to 84 MeV and a converter thickness of up to 0.3 mm. However, the work that still needs to be done is the following: recheck the magnet alignment; estimate the generation of background gamma radiation in the collision of scattered electrons with the magnet; integrate the magnet with the irradiation device.

## CONCLUSIONS

The results of calculations, numerical modeling, design, and testing of a magnetic system to obtain a "pure" beam of bremsstrahlung quanta are presented. This system is made on the base of commercially available permanent magnets with a rectangular cross-section and is designed to study the cross-sections of multiparticle photonuclear reactions at the LUE-40 linac output. The maximum on axis field is 0.9 T, which provides sufficient separation of the electron beam and gamma rays at a distance of more than 90 mm from the magnet.

The developed magnet, in principle, can be used to clean the gamma beam at an electron beam energy of up to 84 MeV and a converter thickness of up to 0.3 mm.

Due to the use of thin targets-converters, the shape of the bremsstrahlung spectrum of gamma quanta is close to the analytical dependence of $1/E_\gamma$. This makes it possible to apply the photon difference method or the regularization method and obtain cross-sections of multiparticle photonuclear nuclear reactions $\sigma(E_\gamma)$.

## ACKNOWLEDGMENTS

The authors thank V.A. Bovda and A.M. Bovda from the Institute of Solid State Physics, Materials Science and Technology NSC KIPT for assistance and useful advice in the design and manufacturing of magnetic systems.